# Multifold enhancement in magnetization of atomically thin Cobalt Telluride

*Solomon Demiss[1,2,§], Raphael Tromer[3,§], Saif Siddique[1], Cristiano F. Woellner[4], Olu Emmanuel Femi[2], Mithun Palit[5], Ajit K. Roy[6], Prafull Pandey[7], Douglas S. Galvao[3,8]\*, Partha Kumbhakar[1]\*, Chandra Sekhar Tiwary[1]\**

[1]Department of Metallurgical and Materials Engineering, Indian Institute of Technology Kharagpur, West Bengal, India
[2]Materials Science and Engineering, Jimma Institute of Technology, Jimma University, Jimma, Ethiopia
[3]Applied Physics Department, University of Campinas, Brazil
[4]Physics Department, Federal University of Parana, UFPR, Curitiba, PR, 81531-980, Brazil
[5]Defence Metallurgical Research Laboratory, Hyderabad, India
[6]Materials and Manufacturing Directorate, Air Force Research Laboratory, Wright Patterson AFB, OH 45433-7718, United States
[7]Department of Materials Engineering, Indian Institute of Science, Bangalore 560012, India
[8]Center for Computational Engineering and Sciences, State University of Campinas, Campinas, SP, 13083-970, Brazil

\*Corresponding Authors E-mail: parthakumbhakar2@gmail.com, galvao@ifi.unicamp.br, chandra.tiwary@metal.iitkgp.ac.in

**ABSTRACT**: Magnetism in semiconductor two-dimensional (2D) materials is gaining popularity due to its potential application in memory devices, sensors, spintronic and biomedical applications. Here, 2D Cobalt Telluride (CoTe) has been synthesized from its bulk crystals using a simple and scalable liquid-phase exfoliation method. The atomically thin CoTe shows over four hundred times enhancement in its magnetic saturation values compared to the bulk form. The UV-Vis absorption spectra reveal superior absorption in the high energy region, suggesting a semiconducting nature. Furthermore, we explain bandgap and origin of high magnetic behavior by density functional theory (DFT) calculations. The 2D CoTe shows a larger magnetism compared to bulk CoTe due to the reduced coordination number of the surface atoms, shape anisotropy and surface charge effect.

## I. INTRODUCTION

Graphene and other two-dimensional materials have recently taken the center stage in research due to their extraordinary physical properties, and have found applications in electronic devices, sensors, biomedical, environmental, energy storage devices, etc.[1–3] Interestingly, magnetism is one property that has not been fully exploited in these materials, even though it had been theoretically predicted back in 1944.[4] The past studies have demonstrated the possibility of introducing magnetic properties in a non-magnetic 2D materials using defect engineering, surface functionalization, doping, and straining.[3,5] However, the magnetism induced by these methods is often weak and short-ranged. Intrinsic magnetism in 2D materials was observed for the first time in 2016 in $Cr_2Ge_2Te_6$ and $CrI_3$.[6] Following this, other 2D materials exhibiting magnetism like $VX_2$ (X= Se, Te), $MnSe_2$, $Fe_3GeTe_2$, $FePS_3$, and FeTe were synthesized, most of them are metallic.[5,7–10]

Magnetic nanomaterials have applications in medical science, electronic device application, water purification, data storage, and also in spintronic devices. Because of recent discoveries of intrinsic magnetism in 2D materials, they are also being considered as potential candidates for these applications. Moreover, 2D magnets will allow the easy fabrication of van der Waals (vdW) materials-based heterostructures with desired physical and chemical properties. For a 2D material to exhibit intrinsic magnetism, it is important for them to have a highly anisotropic electronic structure. It is because of this requirement that in the past, strain engineering, introducing defects, doping, and grain boundary engineering have been employed to induce magnetism in 2D materials.[9,11–13] The intrinsic vdW magnetic materials, on the other hand, have an intrinsic magneto-crystalline anisotropy due to reduced symmetry in their layered structures. This is the reason that even in some of the bulk vdW crystals, effective magnetic interactions can be observed. Even still, ferromagnetism in most of these materials has been limited to low temperatures.

Recently, various transitional metal tellurides are being explored due to their structural, electronic and electrical properties, and magnetic behavior. CoTe has a hexagonal crystal structure of tellurium with the cobalt atoms occupying the octahedral voids.[14] Thus, it is possible to obtain 2D layers of the material. $CoTe_x$ single crystals possess ferrimagnetic behavior having a saturation magnetization of 7.52 gauss/g (0.25 $\mu_B$/Co) and a Curie temperature of 1003 °C in the composition range of $1.00 \leq x \leq 1.20$.[14] More recently, many forms of CoTe, with a variety of nanostructured morphologies, have been synthesized using various physical and chemical methods, such as pulsed laser deposition,[15] hydrothermal synthesis,[16] and chemical transformation route.[17] Magnetic studies of CoTe nanotubes reported a ferromagnetic behavior at 2K with saturation magnetization ($M_s$) of about 13.4 emu/g and the magnetic coercivity ($H_c$) of about 283 Oe.[18]

In this study, we have synthesized high quality atomically thin 2D CoTe from its bulk form using the ultrasonication-assisted liquid exfoliation method. While obtaining 2D materials using liquid-phase exfoliation is not something new and has already been achieved for many materials,[19] they mostly have layered structures with vdW interactions between the layers, making their exfoliation simple. However, recently 2D materials from non-layered (non-vdW) structures have been obtained,[20] which opened up possibilities to obtain a large number of 2D structures. CoTe is also a non-vdW solid and to the best of our knowledge, the present work is the first report of atomically thin 2D CoTe obtained using liquid exfoliation. X-ray diffraction studies confirm the hexagonal crystal structure of CoTe. Scanning electron microscopy (SEM), atomic force microscopy (AFM), and transmission electron microscopy (TEM) were also employed to study the morphology of the exfoliated material. The integrity of the 2D CoTe is analyzed using X-ray photoelectron spectroscopy (XPS) and Raman spectroscopy. We use UV-Vis absorption spectroscopy and Tauc plots to get the optical response and to estimate the bandgap of 2D CoTe, which at ~2.5 eV is slightly smaller than that of the bulk. The room temperature magnetic properties of the 2D CoTe have been measured and compared with its bulk form. To gain further insights on the structural and electronic properties we carried out *ab-initio* density functional theory (DFT) simulation and the obtained results are correlated to the experimental findings.

## II. METHODS

*Synthesis:* Bulk cobalt telluride was obtained via the vacuum induction melting route. Stoichiometric amounts of 99.99% pure Cobalt and Tellurium were taken based on Co-Te binary alloy phase diagram. The alloy was prepared by melting the constituent elements at a temperature of 950 °C in a quartz tube using a vacuum induction melting furnace. Although the melting point of cobalt is 1495 °C, for proper mixing, the furnace temperature should have been higher. But due to the lower boiling point of tellurium (988 °C), a higher temperature would lead to its evaporation and thus, the maximum temperature was limited to 950 °C. Instead, the mixture was kept at an elevated temperature for a long duration to ensure mixing by diffusion. High vacuum conditions of $1 \times 10^{-5}$ mbar were maintained inside the melting chamber to prevent the melt from oxidation. The high vacuum also resulted in lowering of the melting points of the metals, thus ensuring that mixing is proper. The as-cast sample was polished and made ready for further characterization. 2D CoTe samples were prepared using an ultrasonication-assisted liquid exfoliation method. The as-cast sample was powdered by using a mortar and pestle. 50 mg of the powdered sample was dispersed in 150 mL of 2-propanol (iso-propyl alcohol or IPA) and exfoliated in a probe-sonicator for 4h at room temperature to obtain a suspension of 2D CoTe sheets. For the characterization of 2D sheets, we drop-cast the suspension on silicon or glass substrates.

*Characterizations:* The morphology of the samples was observed using a JEOL (JED 2300) scanning electron microscope under a high vacuum with an accelerating voltage of 20kV and 7.475 nA. Energy Dispersive X-Ray Spectroscopy (EDX) was used to determine the chemical composition of the product. The low-resolution TEM and HRTEM images were obtained from an advanced Titan transmission electron microscope. The crystal structure and purity of bulk and 2D CoTe were examined by X-ray powder diffraction (XRD) using a Bruker D8 advance diffractometer with Cu-K$_\alpha$ ($\lambda$ = 1.5406 Å) radiation. For the XRD pattern of the bulk sample, we use powdered CoTe, while for the 2D sheets, we use a drop-casted sample on glass substrate. The scanning range for the samples was 10° - 90° with a step of 0.3° under 40 kV and 100 mA. XRD patterns of both bulk and 2D CoTe samples were analyzed and indexed using Panalytical X'pert high score plus software and Pearson's Crystal Database Software. The surface composition and the oxidation state of exfoliated CoTe were studied using X-ray photoelectron spectroscopy (XPS) (Instrument: ThermoFisher Scientific Nexsa) using the Al-K$\alpha$ radiation (1486.71 eV) as the source of the X-rays. Room temperature Raman Spectroscopy was done using WITec UHTS Raman Spectrometer (UHTS 300 VIS, Germany) at a laser excitation wavelength of 532 nm. UV-Visible Spectroscopy was used to study the optical absorbance and to calculate the optical band gap of 2D CoTe sample.

The measurement of magnetic properties of the bulk and 2D CoTe were performed by employing a vibrating sample magnetometer based on a superconducting quantum interference device (Cryogen-Free Magnet System, Cryogenic Limited, UK). Magnetic hysteresis loops were obtained at room temperature in an applied magnetic field -15 to +15 kOe. Powder sample weight of 62.07 mg CoTe was used to measure the magnetic moment, magnetic saturation, coercivity, and retentivity.

*Theoretical calculations*: Although 2D CoTe undergoes cleavage from the bulk in many crystal directions during sonication, for the theoretical calculations, the structures cleaved along the [100] crystal direction were considered. This choice was based on the fact that the largest x-ray intensity peak was for planes along this direction. The crystal visualization software, VESTA,[21] was used to generate the 2D CoTe structures from the bulk CoTe.

The calculations of the electronic, magnetic, and optical properties of 2D CoTe monolayer were performed with density functional theory (DFT) methods assuming the generalized gradient approximation (GGA) and Perdew-Burke-Ernzerhof (PBE) functional for the exchange-correlation part.[22] We use a mesh cutoff energy value of 300 Ry and 5×5×1 CoTe supercell. A sampling set of the k point was selected with the Monkhorst-Pack scheme.[23] The DFT packages included in the SIESTA software [24,25] were used for the theoretical analysis. Spin polarization was considered for all calculations in this work. The calculated lattice constants for bulk CoTe are a = b = 4.51 Å, c = 20 Å (vacuum region), α = 90.33°, β = 90.33° and γ = 61.21°.

To estimate the absorption optical coefficient, first, we obtained the real $\varepsilon_1$ and imaginary $\varepsilon_2$, components of dielectric function by Kramers-Kronig and Fermi's golden rule, respectively. The absorption coefficient can be calculated by real and imaginary components of the dielectric constant by:

$$\alpha(\omega) = \sqrt{2}\omega[(\varepsilon_1(\omega)^2 + \varepsilon_2(\omega)^2)^{1/2} - \varepsilon_1(\omega)]^{1/2} \quad \ldots (1)$$

It is well known that GGA-DFT methods underestimate the bandgap energy value of materials because the Kohn-Sham equations consider only the ground state charge density. Besides, our system contains Co atoms, composed by valence atomic orbital d, producing large correlation effects. In this work, we adopt the scheme based on the Hubbard model, the so-called LDA+U [26,27] which considers a free parameter that can produce a better description for the band gap value. In order to fit this parameter, we consider the same strategy from Ref. [20] and choose the Hubbard parameter which best reproduces the experimental band gap values.

## III. RESULTS AND DISCUSSIONS

**Figure 1a** depicts the crystal structure and schematic of the exfoliation of bulk CoTe into 2D CoTe using a liquid exfoliation method. The digital photograph shows the bulk CoTe crystal and the exfoliated CoTe (right), which is dispersed in isopropyl alcohol (IPA). The observed red laser light in the exfoliated CoTe confirms a good dispersion of 2D CoTe sheets. Bulk CoTe has a NiAs-type hexagonal structure where Co atoms are in octahedral coordination, surrounded by six Te atoms. Te atoms are also coordinated to six Co atoms and occupy the trigonal prismatic sites. Because of the hexagonal lattice of the CoTe, it is possible to exfoliate it and obtain 2D layers. Top and side views of the obtained [001] oriented 2D CoTe illustrate that the bulk CoTe is exfoliated into hexagonal crystallographic planes oriented in (001) plane. However, 2D CoTe presents a buckling close to 0.8 Å. The optimized parameters of the hexagonal unit cell, obtained by DFT calculations, considering a vacuum region of 20 Å were: a = b = 4.51 Å, and $\gamma$ = 61.21° and these values are close to the experimental ones. We have found that the formation energy for the 2D CoTe sheet is 2.97 eV/atom. The lower formation energy indicates thermodynamic stability and the feasibility of extraction from bulk. For a better understanding of the structural properties of the 2D CoTe structure, we carried out a set of first-principles calculations using the SIESTA package.[24] As mentioned before, we considered the structure cleaved from bulk along the [100] crystal direction because the peak corresponding to it showed the largest X-ray intensity. **Figure 1b** shows the geometrical optimization of one (1L) and four-layer (4L) for the 2D CoTe. For 1L we extracted the initial configuration from bulk and performed the optimization process considering that atoms and lattice vectors are optimized simultaneously, and the convergence is achieved when the forces on each atom were less than 0.05 eV/Å. For more than 1L we considered the same optimized lattice vectors values from 1L and re-optimized only the atoms from the initial configuration extracted from bulk. This ensures that both 1L and multilayer have lattice vectors close to the experimental values. The optimal structure has a bond length of about 2.5 Å between Co and Te atoms.

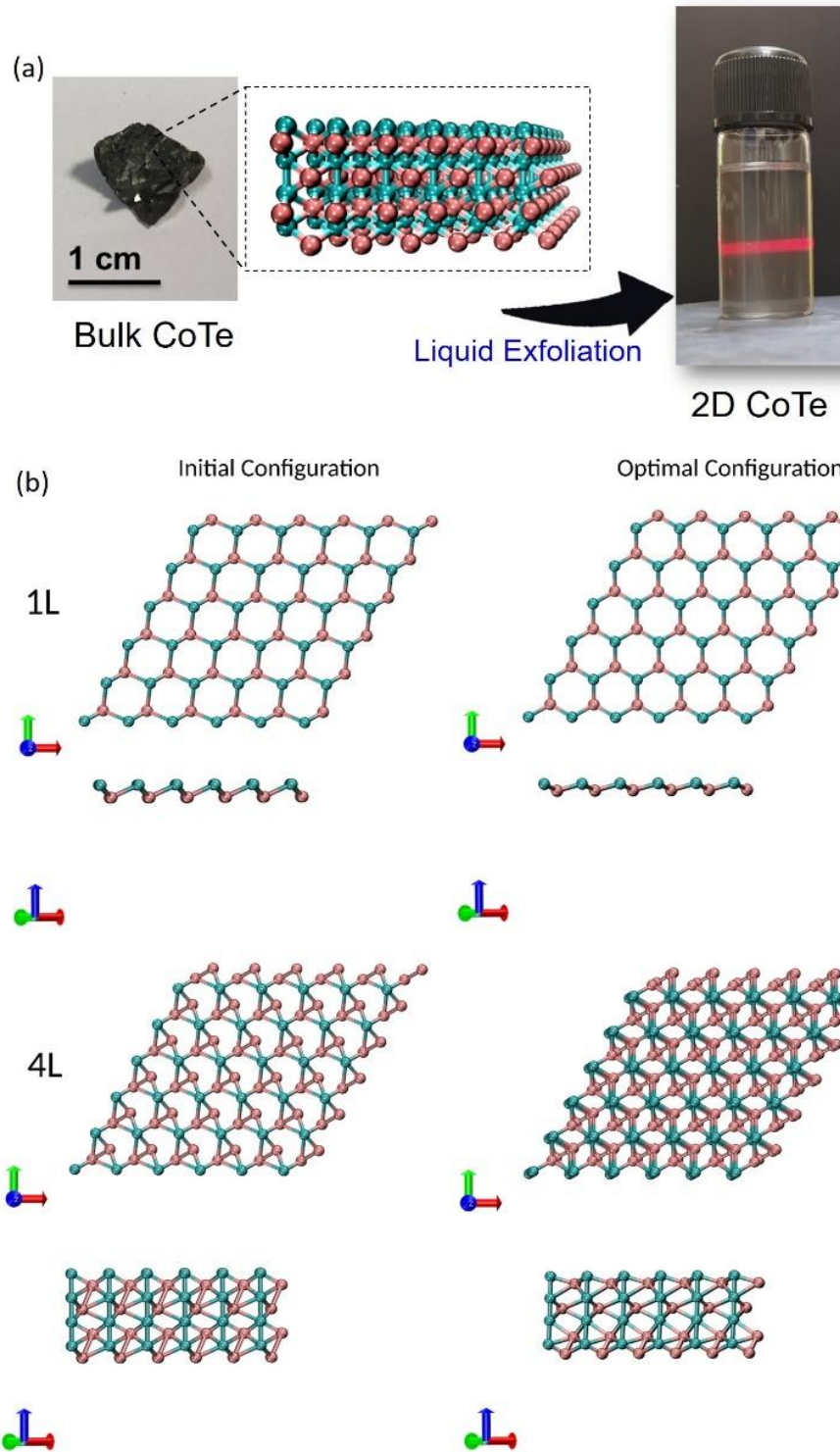

**Figure 1.** (a) Schematic representation of synthesis mechanism of 2D CoTe from bulk CoTe crystal (at left) in IPA with digital photograph of liquid exfoliated CoTe (at right). Cobalt and tellurium atoms are colored red and green, respectively, (b) top and side views before and after the optimization process for one (1L) and four layers (4L). We can see that the optimization process produces untwisted rings. The buckling height for one layer is about 0.84 Å.

Bulk and exfoliated CoTe were characterized by Scanning Electron Microscopy (SEM). **Figure S1** shows the SEM image of bulk CoTe clearly revealing the micrometer-sized large grains with multilayer stacking morphology. Well-separated layers of CoTe can be seen from the SEM image of exfoliated CoTe (**Figure 2a**). The TEM image in **Figure 2b** also depicts the ultra-thin nature of the exfoliated 2D CoTe. **Figure 2c** shows the HRTEM image of 2D CoTe. Inset shows the lattice spacing and it was measured as ~0.32 nm corresponds to (100) plane. The lower inset shows the FFT pattern corresponding to a square box of the HRTEM image. An elemental analysis performed using Energy Dispersive X-Ray Spectroscopy (EDX) (**Figure 2d**) reveals that the obtained material consists only of Co and Te elements with nearly the same atomic percentage ratio, which is consistent with the expected stoichiometric ratio. We employed atomic force microscopy (AFM) to measure the thickness of the 2D CoTe. The AFM image confirms the formation of 2D thin sheets, as shown in **Figure 2e**. The height of the sheet is ~2 nm measured from the AFM scan (**Figure 2f**) which demonstrates the presence of 3-6 atomic layers in the exfoliated material.

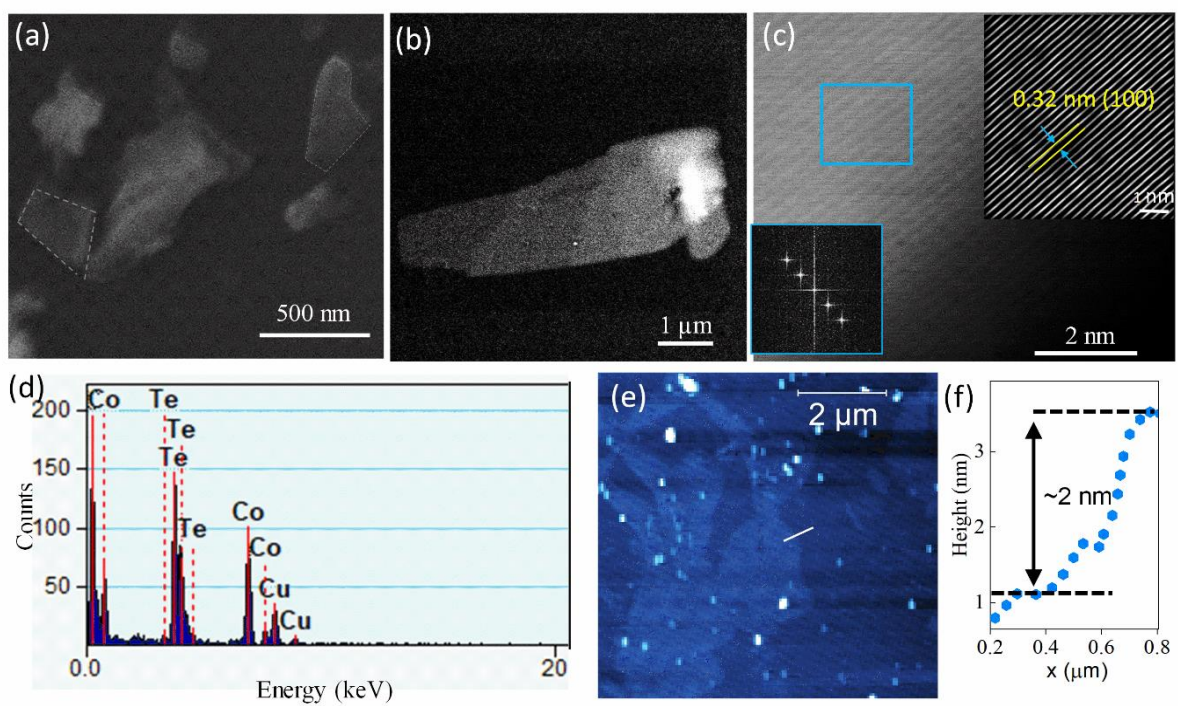

**Figure 2.** (a) SEM image of exfoliated few-layer CoTe sheets. (b) Low magnification TEM image of exfoliated 2D CoTe. (c) HRTEM image of ultrathin CoTe. Inset shows the (upper) lattice plane of 2D CoTe. Lower inset shows the FFT pattern. (d) EDX spectra of CoTe. (e) AFM image of ultrathin CoTe. (f) The height profile of 2D CoTe.

The X-ray diffraction (XRD) results of the bulk and 2D CoTe are shown in **Figure 3a** and are indexed in the hexagonal crystal structure, which belongs to the space group *P63/mmc*. We obtain lattice parameters $a = b = 3.89$ Å, and $c = 5.36$ Å, which are in good agreement with previously reported results.[15] For both bulk and exfoliated samples, all the diffraction peaks correspond to (100), (101), (002), (102), (200), (201), and (112) planes of the hexagonal phase. Among these peaks, the (100), (101), and (002) diffraction peaks are dominant in 2D CoTe, which indicates that compared to others, these planes were exfoliated in excess during liquid exfoliation from bulk. The observed difference in peak intensity can be due to the different crystallographic orientations in bulk and 2D sheets. As shown in **Figure 3a**, no extra peak was observed after liquid exfoliation, indicating good stability and purity of the exfoliated CoTe. The presence of lattice strain ($\varepsilon$) is calculated using Williamson Hall plot and it is presented in **Figure 3b** (see supporting information for details). It has been observed that the 2D CoTe has more strain compared to that of bulk CoTe and it clearly suggests that the formation of surface defects in the 2D surface is due to exfoliation. Raman spectroscopy was used to confirm the successful synthesis of 2D CoTe, and the measurements conducted on drop-cast samples with a laser excitation of 532 nm. For exfoliated sample (**Figure 3c**), five prominent Raman peaks at 52, 109, 117, 122, and 145 cm$^{-1}$ were obtained which correspond to the Raman vibrational modes of $A^1_g$, $E^2_g$, $A^2_{1g}$, $A^1_u$ and $E_{T0}$.[28,29] The intensity of Raman modes increased as the thickness decreased. We have also taken the Raman measurement for bulk CoTe, as shown in **Figure 3c**. The intensity of the Raman modes is very low as compared to the of exfoliated CoTe. Furthermore, a small redshift was observed when exfoliating the bulk to 2D CoTe due to weak interactions among the interlayers. The increase or decrease of the Raman shift depends on the different types of phonon modes as well as on the number of layers.[30] X-ray photoelectron spectroscopy (XPS) measurements are employed to elucidate the oxidation state of Co and Te and their atomic ratio (survey spectra, **Figure S2**, supporting information). **Figures 3d-e** illustrate the Cobalt and Tellurium peaks in 2D CoTe with corresponding binding energies and reveals that the sample is composed entirely of Co and Te. **Figure 3d** presents the Co 2$p_{3/2}$ and Co 2$p_{1/2}$ peaks at 780.9 eV, and 797.2eV, respectively. These are presumably attributed to the divalent oxidation state of Co. The peaks at 573.8 eV and 584.3 eV are due to Te 3$d_{5/2}$ and Te 3$d_{3/2}$ of the Te$^{2-}$ ions, respectively (**Figure 3e**). The satellite features for Co 2$p$ at 786.2eV and 803.6 prevails the existence of Co(II).[31] From the XPS studies, Co and Te in our samples are in the molar ratio of 1:1.

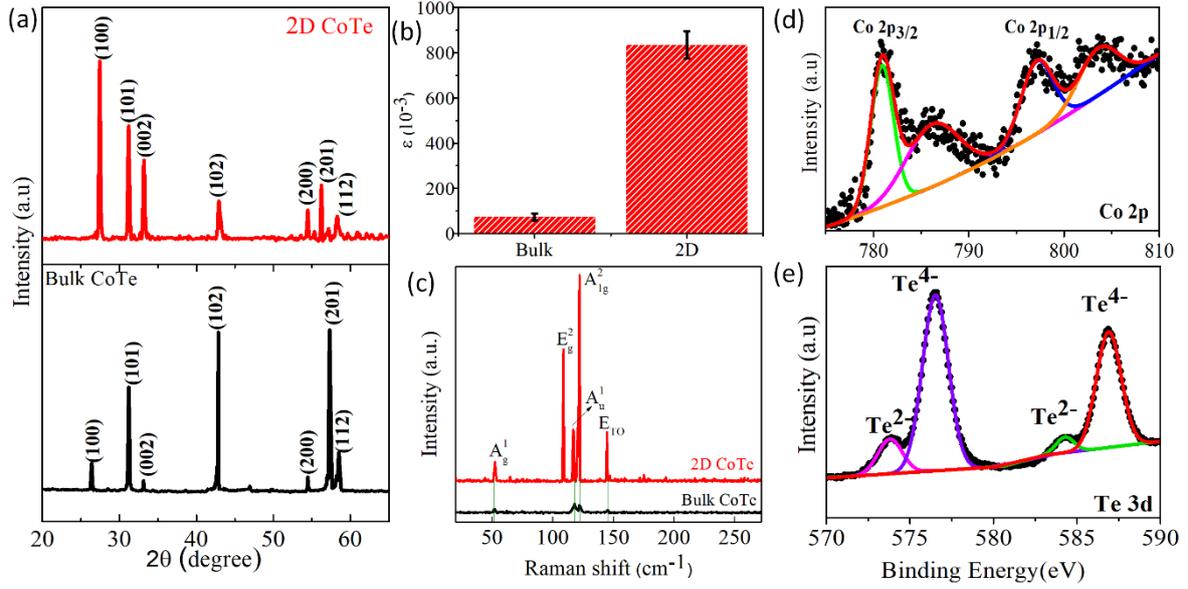

**Figure 3.** (a) XRD patterns of bulk and exfoliated CoTe, (b) Bar chart of calculated strain for both samples. (c) Raman spectra of buk and 2D CoTe. High resolution XPS spectra corresponding to (d) Co2p and (e) Te3d of 2D CoTe

## A. Optical and Electronic Properties of 2D CoTe

The optical performance of 2D CoTe was analyzed using UV-Vis Absorption Spectroscopy in the energy range of 1.5 to 3.5 eV. **Figure 4a** shows the UV-Vis absorption spectrum. We use Tauc plot to estimate the bandgap of 2D CoTe. Inset of **Figure 4a** shows the Tauc plot between $(\alpha h\nu)^{(1/n)}$ vs. $(h\nu)$ where, α is the absorption coefficient, calculated from absorbance, and thickness of the sample, $h\nu$ is the photon energy, and the power factor (*n*) taken as a value of 2 for direct bandgap semiconductor. To determine the energy bandgap value using Tauc plot, we extrapolate the linear region of the UV-Vis absorption spectrum to the zero absorption coefficient value (abscissa). The estimated value of $E_g$ is ~2.5 eV, and is larger than the corresponding values of bulk CoTe. The increase of the bandgap from bulk to 2D is due to the quantum confinement effects. The optical properties of the 2D CoTe monolayer were also examined by using DFT calculations. The calculated absorption coefficient α(ω) in a wide energy range is presented (**Figure 4b**) for the CoTe monolayer. The threshold energy of the absorption spectra occurs at around 1.5 eV and 2.2 eV for the spin up and spin down, respectively. Two maximum peaks for the spin up and spin down are clearly observed in these plots. The first peaks correspond to the direct optical transition for the spin-up and spin down as clearly observed in these plots. For the spin up case, the first peak is situated at 3.3 eV and for the spin-down one, the first peak is observed at 4.0 eV, both of which are in the UV region.

For a material to be suitable for photocatalytic and water splitting activity, it should have an appropriate band-gap value, high carrier mobilities, and absorb a significant portion of the incident UV-visible light. The DFT calculations and experimental results suggest the suitability of 2D CoTe monolayer for catalytic applications in both visible and UV regions.

The electronic band structure of 2D CoTe was calculated using DFT. **Figure 4c** shows the bandgap energy as a function of the Hubbard parameter U. Note that for U = 0 eV (no-correction) the system is a semiconductor with a gap around 0.5 eV. However, this value is significantly smaller than the one determined experimentally, close to 2.5 eV. Bandgap values increase linearly with U until near 5.0 eV and then oscillate until 10 eV. A value of U = 5.0 eV was chosen to correctly obtain the bandgap value close to 2.5 eV. The dynamical stability of 2D CoTe was tested by computing phonon dispersions along the high symmetry Γ-M-X-Γ direction. The absence of negative and imaginary phonon modes (**Figure 4d**) in the phonon dispersion diagram of CoTe monolayer clearly rules out any structural instability of CoTe monolayers. As shown in **Figure 4e**, the electronic band structure shows that CoTe monolayer has a direct bandgap. Both the valence band maximum (VBM) and conduction band minimum (CBM) are situated in the same direction of the first Brillouin Zone. Note that we shifted the Fermi level value $E_F$ to coincide to zero, thus redefining the energy axis as $E-E_F$. **Figure 4f** shows the PDOS (projected density of states) of 2D CoTe monolayer for only the valence atomic orbitals of Te and Co atoms and considering U = 5.0 eV. A more detailed PDOS is shown in **Figure S3**. The vertical dotted line in the figure represents the Fermi level as a reference and set to 0 eV. The atomic orbitals, 5*s* of Te and 3*d* of Co contribute more in PDOS. The Fermi level is located close to the conduction band. The band structure and PDOS results confirms that monolayers of CoTe is a n-type direct wide bandgap semiconductor.

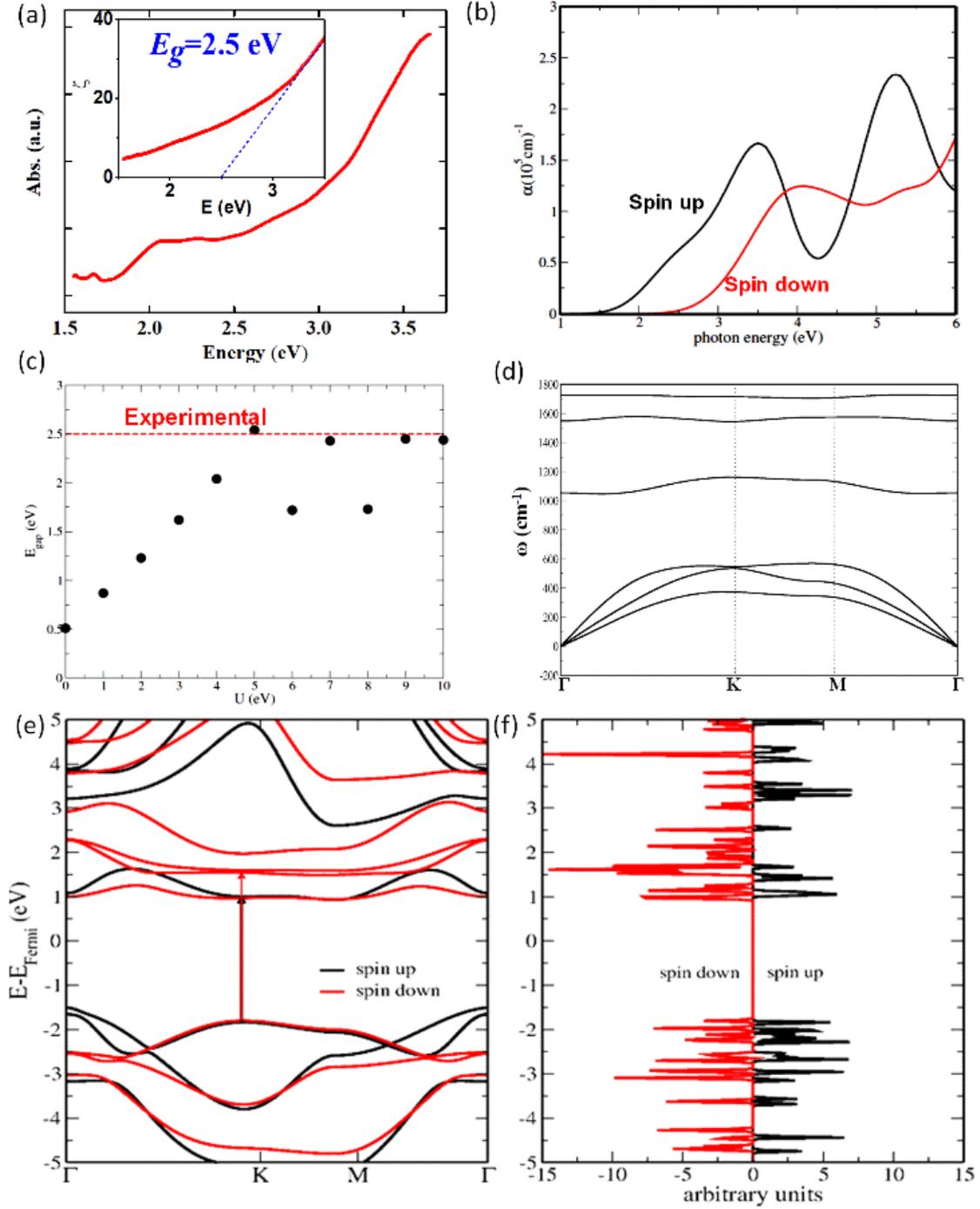

**Figure 4.** (a) UV-Vis absorption spectra of 2D CoTe sheets (Inset shows the Tauc plot of 2D CoTe sample). (b) Spin contribution for absorption coefficient as a function of photon energy. (c) Gap energy as a function of the Hubbard parameter U. (d) Phonon dispersion in CoTe monolayers. (e) Band structures of 2D CoTe monolayer with Hubbard parameter U=5.0 eV. the dotted line represents the Fermi level. (f) Projected density of states (PDOS) for spin up (majority) and spin down (minority) for the valence atomic orbitals of Te and Co atoms considering U = 5.0 eV. As standard, the Fermi level was set to 0 eV.

**B. Magnetic Behavior of 2D CoTe**

The room temperature (300K) magnetic behavior of the as-synthesized bulk and atomically-thin CoTe was investigated using Vibrating Sample Magnetometry, as shown in **Figure 5a**. The *M-H* curves of both bulk and 2D CoTe show magnetic ordering. However, the magnetic ordering of the bulk CoTe is very weak at room temperature as can be seen from the *M-H* curve shown in **Figure 5a**. This type of extremely weak magnetism has been previously observed.[14,32] Saturation magnetization ($M_s$) of about 0.0908 emu/g (or 0.003 $\mu_B$/Co atom, where $\mu_B$ is Bohr Magneton) and magnetic coercivity ($H_c$) of about 45.66 Oe (3.63 x $10^3$ A/m) was observed for bulk CoTe. However, at 300K, $M_s$ of 2D CoTe is 37.38 emu/g (~1.25$\mu_B$/Co atom), which ~400 times higher than bulk CoTe (**Figure 5b**). This value is almost three times larger than those reported for CoTe nanotubes at 2K.[18] The low magnetic saturation in bulk CoTe is due to the antiparallel coupling of Co moments in the different hexagonal layers.[33] This might also be the reason why a high magnetisation has not been achieved in other CoTe nanostructures. However, in monolayer and few-layer 2D CoTe, the Co moments from different layers do not cancel each other and we get a high magnetic saturation value.

When ferromagnetic materials are converted to ultrathin nanostructures, they become superparamagnetic and show no hysteresis. This type of behavior has been observed in iron oxide nanoparticles and is the result of the presence of single magnetic domains. The magnetic saturation of such materials is higher than those of paramagnetic materials, although the coercivity is zero. However, in the case of exfoliated CoTe, a hysteresis loop with a coercivity of 174.304 Oe (1.39 x $10^4$ A/m) is observed indicating a ferromagnetic/ferrimagnetic nature (**Figure 5c**). The increase in the saturation of the magnetization of the exfoliated material is due to the increase in the overall density of states especially for the Co 3*d* band near the Fermi level. The enhancement of coercivity of the 2D material is due to the localization of Co 3*d* state leading to an increased spin-orbit coupling and thus magneto-crystalline anisotropy. Due to the nature of the exfoliated materials, it is important to take into consideration possible extrinsic factors in the measure magnetisation. Accordingly, in our de-convoluted XPS spectra, we observed the presence of cobalt in its oxidized state, instead of metallic. Therefore, it is possible that the oxidation state of Co can be contributing to the observed magnetism in the 2D CoTe. In most of the Co-based oxides, the magnetic ordering temperature is far below the room temperature (RT), except CoO, which exhibits Néel temperature close to RT (precisely 298 K). The hysteresis loop is extremely sensitive to the presence of even small quantities of CoO. Because of the antiferromagnetic nature of CoO, there is in an asymmetric shift in the hysteresis loop along the applied field axis. However, in the present measurement, no such shift has been noticed at RT measurement and hence, the presence of CoO can be ruled out. Moreover, since all other Co-based oxides are paramagnetic at RT, their contribution towards magnetization can be ignored as the observed value scales much higher. The enhancement of coercivity in the 2D CoTe is primarily arising from larger spin-orbit coupling of the localized electrons in the spin down 3$d$ band of Co (**Figure S3**). Further, 2D CoTe also possess shape anisotropy due to the formation of thin layers which also leads to the enhancement of coercivity. **Figure 5d** shows a map of the different 2D magnetic chalcogenides: their RT bandgap and magnetic susceptibility.[10,34–36] It can be clearly observed that 2D CoTe has a much better performance as a semiconductor as well as a ferromagnetic material.

DFT calculations show (**Figure 5e-f**) that the system presents a magnetic character at T = 0 K because of the difference in the populations between spin-up and down. We confirm the magnetic order at T = 0 K from energy difference between ferromagnetic and antiferromagnetic states, where $\Delta E = E_{total}(ferro) - E_{total}(antiferro) = -2.2$ eV. It indicates that the 2D CoTe is stable at T = 0 K with a ferromagnetic order. The total magnetic moment for CoTe monolayer, $\vec{\mu}$ = 3.0 $\mu_B$, is larger than the reported value for bulk CoTe. This value of magnetic moment is also larger than our experimental value of ~1.25 $\mu_B$. This is because of absence of proper alignment of the moments among the different layers of the few-layers CoTe in the exfoliated dispersion. **Figure S4(b)** shows the variation in magnetic moment as the number of layers are increased. The enhancement of magnetic moment in 2D CoTe can be explained based on the well-known Stoner-like model of itinerant magnetism.[37,38] When bulk CoTe is exfoliated into a 2D CoTe monolayer, the electronic band structure changes from its bulk counterpart. The 3$d$ band of the Co atom becomes narrower near the surface as a result of the reduced coordination number of the surface atoms. This narrow energy band enhances the exchange splitting and magnetic moment in 2D CoTe. In short, the magnetic moment per atom near the surface will become larger than its value at the bulk.

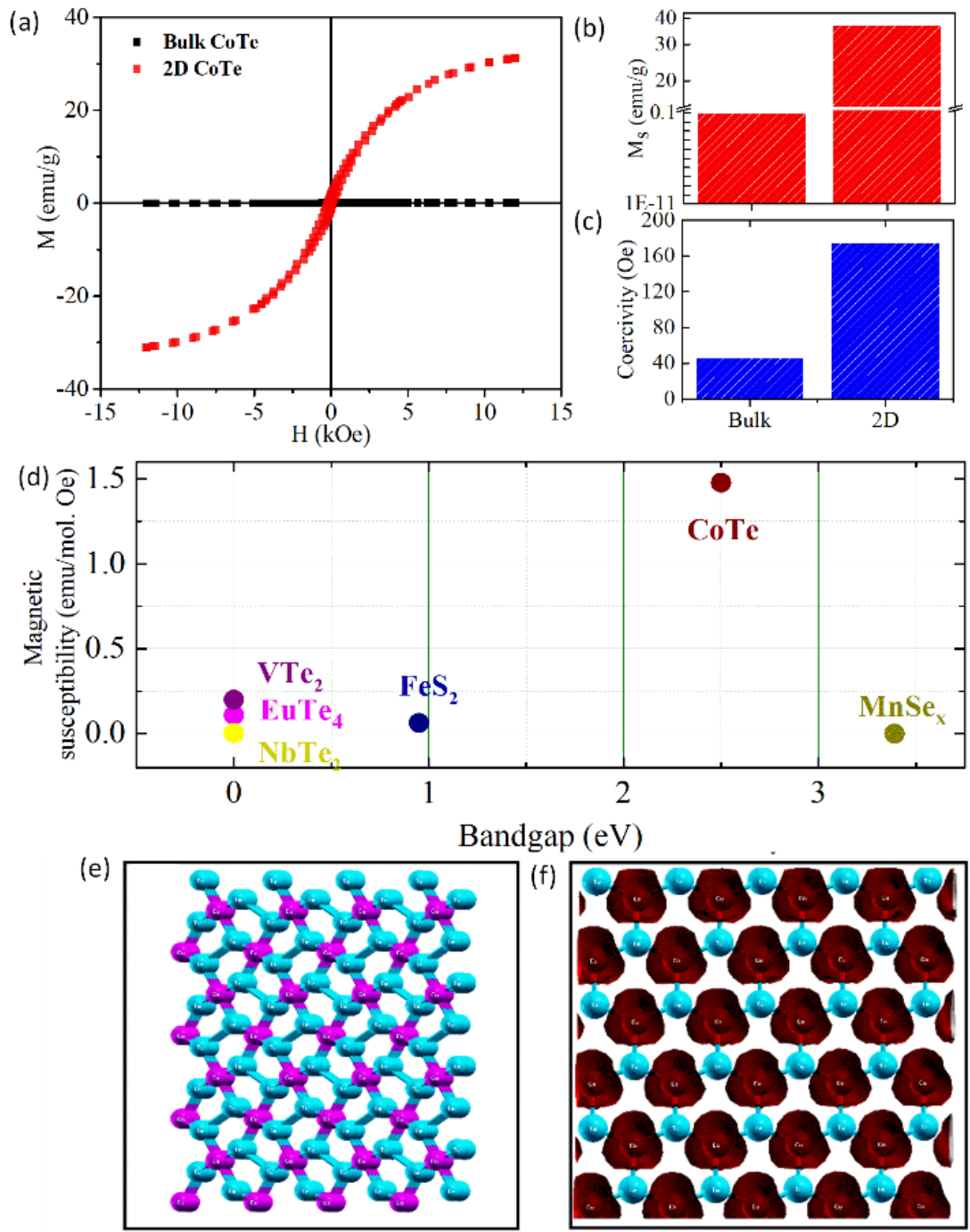

**Figure 5**. Magnetic measurements of CoTe: (a) M–H curves of 2D and bulk CoTe at room temperature (300K). (b) bar chart representation of magnetic saturation. (c) bar chart representation of coercivity of 2D CoTe. (d) A comparison between different 2D magnets.[10,34–36] The charge density difference plot for the spin up and down cases in (e) bulk and (f) 2D CoTe.

In **Figures 5e** and **5f**, we presented the charge density difference plot for the spin up and down cases, for the bulk and the 2D CoTe, respectively. We can see that for the bulk case in the **Figure 5e**, the spin population charge difference is null, that is, the charge quantity presented in the orbitals are the same for the spin up and down cases. In contrast, for the 2D CoTe case showing in **Figure 5f**, there exists a liquid charge localized in the Co atoms giving rise to the high magnetic moment and explaining the different magnetic behavior of the bulk and 2D systems.

We also examined other multilayer structures corresponding to 2L, 3L, and 4L, assuming the same parameters used for 1L, where the used Hubbard parameter was U = 5.0 eV. We observed that one-layer has the largest values for the gap energy and magnetization. More importantly, some characteristics of the 1L, such as the ferrimagnetic order and the semiconductor character are preserved for multilayers (but not for the bulk). **Figures S4(a)** and **S4(b)** in the supplementary materials present the bandgap and moments as a function of the number of layers. Although the multilayers exhibit a bandgap value smaller than the experimental one, the discussion when contrasted to the bulk is similar to one-layer case. For multilayers we observe that the magnetic moments, and in turn the magnetization values are always larger than that of the bulk. The magnetic moment decreases from the maximum value of $3.0\mu_B$ (for 1L) as the number of layers are increased. **Figure S5** (in the supplementary material) shows the charge difference plot for spin up minus spin down for 4L. We can see that some Co atoms do not contribute to magnetization because the net charge is zero. Thus, the values should be smaller than 1L, where all Co atoms have a net charge different from zero. In the bulk limit, the magnetization is null because the net charge is again zero for all Co atoms, as we can see from **Figure 5f.**

## IV. CONCLUSIONS

Few-layer CoTe were obtained by ultrasonication assisted liquid exfoliation route. The obtained bandgap of CoTe indicates that exfoliated sample has a semiconducting nature. Magnetic characterization using VSM reveals a ferromagnetic behavior in 2D CoTe even at room temperatures, where other nanostructures achieve the same at near absolute zero temperature, which is further confirmed using the DFT calculation. In 2D CoTe, the 3*d* band of the Co atom becomes narrower which enhances the exchange splitting and magnetic moment per atom near the surface. Additionally, localization of surface charge is giving rise to the high magnetic moment and explains the different magnetic behavior of the bulk and 2D systems. Therefore, the coexistence of 2D magnetic and semiconducting behavior, that too at room temperature, makes 2D CoTe an ideal candidate for spintronics and also opens up the opportunity to form heterostructures with other van der Waals materials.

**AUTHOR'S CONTRIBUTIONS**

§-SD and RT contributed equally.

**ACKNOWLEDGEMENTS**

SD thanks Jimma Institute Technology, Jimma University and Ministry of Science and Higher Education Ethiopia for their funding support. We also thank Indian Institute of Technology Kharagpur for laboratory and characterization facility support. This work was financed in part by the Coordenacão de Aperfeiçoa- mento de Pessoal de Nível Superior - Brasil (CAPES) - Finance Code 001, CNPq, Brazil, and FAPESP, Brazil. RT, CFW, and DSG thank the Center for Computational Engineering & Sciences (CCES), Brazil at Unicamp for financial support through the FAPESP/CEPID Grant 2013/08293-7.